\documentclass{IEEEcsmag}

\usepackage[colorlinks,urlcolor=blue,linkcolor=blue,citecolor=blue]{hyperref}
\expandafter\def\expandafter\UrlBreaks\expandafter{\UrlBreaks\do\/\do\*\do\-\do\~\do\'\do\"\do\-}

\jvol{XX}
\jnum{XX}
\paper{8}
\jmonth{July}
\jname{IEEE Internet Computing}
\jtitle{Consumer-Oriented Computing: A Path to Community Data Centers}
\pubyear{2025}

\begin{document}

\sptitle{Spotlight}

\title{Consumer-Oriented Computing: A Path to Community Data Centers}

\author{Tianhao Zhang}
\affil{Independent Researcher}

\markboth{SPOTLIGHT}{SPOTLIGHT}

\begin{abstract}\looseness-1
Modern large-scale data centers are known for their engineering complexity, cooling, and oversubscription challenges. To mitigate these issues, this article proposes the implementation of community data centers that are closer to consumers as part of the data center ecosystem. Having a community data center can reduce latency, minimize network burden on Internet Service Providers (ISPs), utilize full computing capability, available during disaster events, and simplify the engineering complexity associated with traditional data centers. In addition to that, this article explores one technical design for such a community data center and the business strategy for operating community data centers.
\end{abstract}

\maketitle

\chapteri{D}ata centers have the impression of large, noisy data halls with thousands of machines running. They consume a lot of electricity and generate immense heat from the servers. According to a report by the US Department of Energy, data centers can consume 10 to 50 times the energy compared to a typical commercial office building and contribute to 2\% of the total U.S. electricity use\cite{DOE}. With a large number of servers that consume energy, cooling can be a major challenge. To remedy this, some measures are taken to keep the data center underwater\cite{Natick}, and others are implementing measures to redirect heat to nearby residential homes, greenhouses, and farms\cite{AzureHeat}. 

Within data centers, there is engineering complexity to manage servers at scale. On the Amazon EC2 page, there are 106 instance types listed, categorized into General Purpose, Compute Optimized, Memory Optimized, Accelerated Computing, Storage Optimized, and HPC Optimized instances\cite{AWSExplorer}. These varieties of servers represent software and hardware engineering to accommodate servers of different shapes and sizes. Especially with the rising popularity of GPU servers due to AI demands, cloud vendors will need to accommodate power-hungry GPUs, which can create strains on data center power and thermal capacity. Statistics show that emerging GPUs are rated at up to 700W TDP, more than twice as much as a large server CPU, and each GPU server can house 4 to 16 GPUs\cite{Patel2023}.

As a result, data centers use oversubscription techniques and throttle the GPU servers to maintain cooling/power requirements. This is implemented via in-band and out-of-band management. As an example, Nvidia’s SMI and DCGM tools provide in-band capability to throttle power on Nvidia GPUs, while Nvidia’s proprietary SMBPBI interface allows the power cap to be set out of band\cite{Patel2023}. The use of power throttling techniques could lead to a loss of peak computing potential, which directly corresponds to wasted computational resources.

\section{The Consumer Computing Demand}
\vspace*{5pt}
Today, data centers are not used directly by average consumers. Based on Oracle\cite{OracleStories}, Google\cite{GoogleCloud}, and Amazon\cite{AWSCustomers} customer pages, they are ubiquitously used by enterprises, government \& public sectors, and research facilities. Although average consumers do not need as much computing power compared to enterprise customers, there is a trend that gaming and AI are taking a bigger role in terms of computational resources. 

\subsection{Consumer Demands on Gaming}
In 2024, approximately 61\% of Americans ages 5 to 90 years play video games\cite{GamesStats}, but the cost of computing resources to achieve recommended gaming experiences has increased significantly over the past two decades. As an example, Tomb Raider 2013 recommended a GTX480 (with 1GB graphics RAM)\cite{TombRaider}, whereas Ray Tracing Cyberpunk 2077 recommended a Nvidia RTX 3080TI (with 12GB GDDR6X graphics RAM)\cite{Cyberpunk}. Looking at the MSRP of GTX480 and RTX3080TI, the price doubles from \$499 to \$1199\cite{AnandTech480}\cite{PCMag3080}, which can be a burden for average consumers. In addition, even dedicated gamers will not be able to play games all the time, and there could be a waste of expensive computational resources while gamers are not using their desktops.

\subsection{Consumer Demands on AI}
With the rise of generative AI tools like ChatGPT and DALL-E, consumers are starting to use AI to create text, images, and even music for personal or professional projects. Those tools produce great results for content on the Internet, but they do not provide the consumer with the context based on their personal preference or private data. On 13 February 2024, Nvidia released the “Chat with RTX” tool to allow the general public to connect private user data with the open-source large language model using Retrieval-Augmented Generation (RAG) technology\cite{ChatRTX}. Based on my test using the RTX3060 TI, this worked great on a limited number of small files. However, it would take hours if not days to fine-tune the model with all the documents on my desktop computer. In addition, RTX series graphics cards are intended for creative workloads and gaming instead of AI training, so data center GPUs like Nvidia A100, H100 would suit better for such processing needs. A similar effort by Apple is trending towards providing personalized AI in a private cloud that is subject to verification by independent researchers to ensure security\cite{AppleAI}. Although it is early to accurately assess the demand of this new market, there is a trend of providing personalized AI model to consumers with its characteristics of recurring training (from new private data), usage across multiple devices (i.e. phones, tablets, smart speakers, and computers), and an emphasis on security to avoid private data leak.

\section{Introduction to \\Community Data Centers}
\vspace*{5pt}
To summarize from the above sections, current data centers are complex and may not reach their full computing potential. Furthermore, there is a gap between data centers and average consumers in terms of gaming demand and AI computing needs. 
\\\\
A data center closer to the end consumer is needed. The term “Community Data Center” used in this article refers to edge data centers but with an emphasis on people instead of the proximity between data processing and data generation. In this way, we can start from people’s needs and not technical aspects. The community data centers proposed in this article include small sets of homogeneous GPU servers located near the community gateway to serve the community's gaming and AI needs. 
\\\\
The aforementioned community data centers are intended to be deployed in luxury residential areas, schools, universities, and population-dense areas. Based on results from the 2020 Census, 80.7\% of the US population lives in urban areas\cite{USCensusUrban}. In most densely populated urban areas, such as Los Angeles and New York City, there could be as many as 7400 and 5980 people per square mile\cite{USCensusPop}. Therefore, deploying such data centers in urban areas would fit the need of most US populations.

\section{Benefits and Risks of \\Community Data Centers}
Community data centers offer low latency and reduced network burden by locating servers closer to users, which is ideal for applications like cloud gaming and AI workloads. They avoid power and network oversubscription issues due to their smaller scale, support disaster recovery through external power sources, and benefit from a simplified, cost-efficient design. However, they face challenges in scalability due to space and cost limitations and generally lack the robust physical and software security found in large-scale data centers, posing risks for sensitive AI tasks.

\subsection{Advantages of Community Data Centers}
\subsubsection{Low latency and network burden}
To access the broader Internet, residential users will need to go through multiple ISP network routers and even a partner network. Depending on the network topology, the researchers tested 8098 pairs of Internet sites and reported an average hop count of 17.0\cite{Fei1998}. The more network switches and routers users encounter along the way, the higher the latency users will experience. This is particularly noticeable in cloud gaming, where the graphics are computed in the cloud and streamed to the user's device. By having community data centers, the servers are much closer to the end user. With the growing popularity of the fiber Internet, community data centers will yield even better latency results. Also, with less traffic flowing through the ISP network, the bandwidth for network-intensive cloud gaming and AI model training can be made available to other use cases.

\subsubsection{No oversubscription on power/network resources}
As mentioned in the data center section, traditional data centers have power limitations and struggle to handle a large GPU presence. That is the reason behind the use of power throttling and network oversubscription techniques to reduce load. In community data centers, the number of server racks to support is much smaller, which creates fewer cooling/power strains.

\subsubsection{Disaster recovery}
During a large-scale power or weather event in the area, connectivity to local hyperscale data centers may not be available. A community data center can be supported by external power generators and maintain levels of availability.

\subsubsection{Modular and simplicity design (with homogeneous hardware)}
Community data centers can use homogeneous server hardware to support their focused use cases from an AI and gaming perspective. For example, by only having two types of servers (Nvidia H100 GPU server and RTX 40 series servers with fixed memory / CPU configuration), costs can drop significantly. 

\subsection{Disadvantages of Community Data Centers}
\subsubsection{Scaling}
One of the major drawbacks of a community data center is that the space available to the facility can be expensive and limited. It does not scale quickly when demand surges in a short period of time. To solve this, a tiered data center ecosystem approach and waiting list can be implemented. When the community data center is overwhelmed with gaming requests, the requests will be rerouted to the centralized data center. For AI tasks that are not time-critical, a task queue can be maintained and data will be computed when hardware is available. In this way, the utilization of the GPU hardware stays high.

\subsubsection{Security}
Community data centers are generally small and do not have high security compared to large-scale data centers that house tens or thousands of servers. For gaming racks, data security may not be as important, but AI racks that contain consumer private data require much higher security, both physically and in software. The details of this will be described further in the technical design section below.

\section{Technical Design of \\Community Data Centers}
\begin{figure}
\centerline{\includegraphics[width=18.5pc]{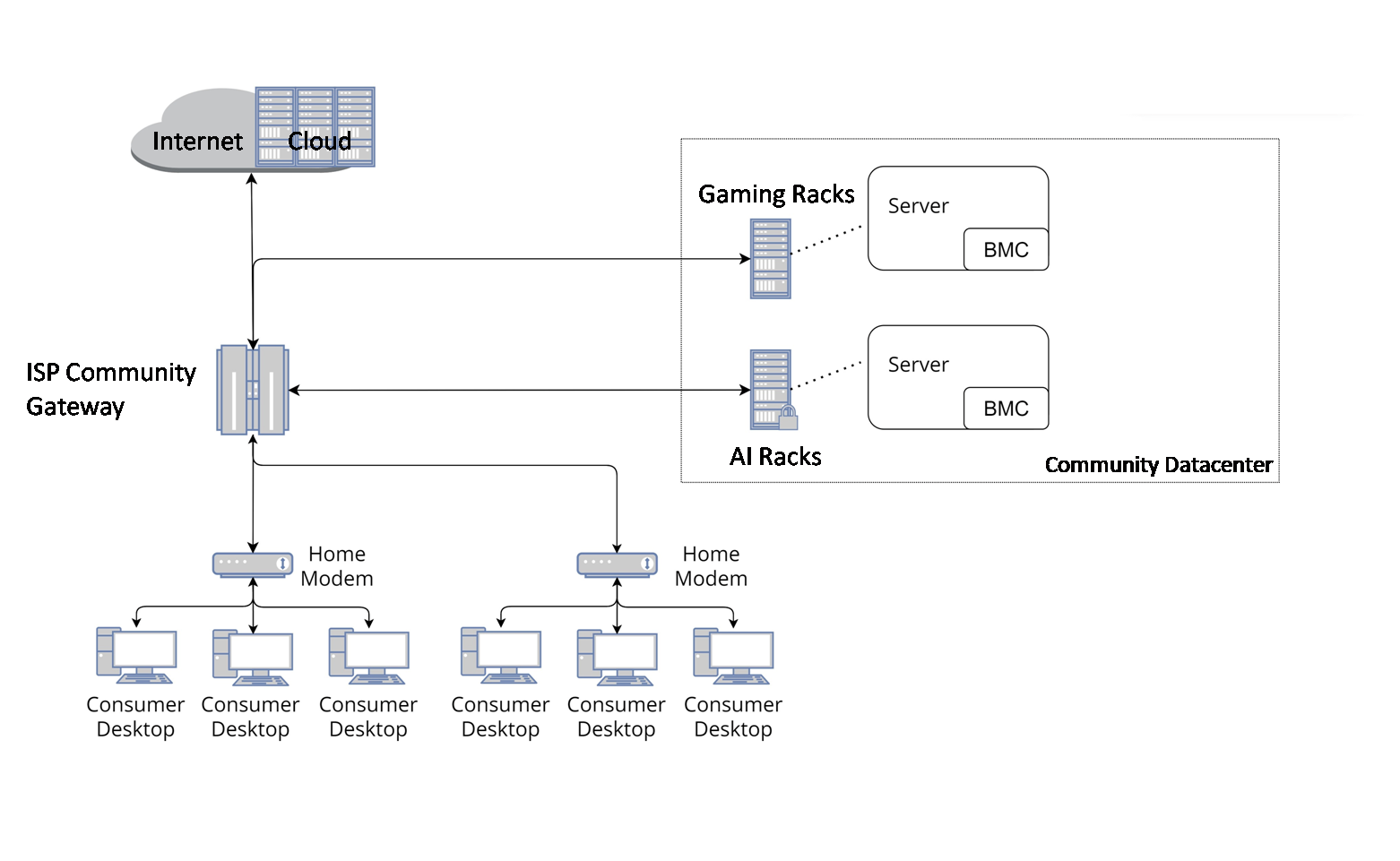}}
\caption{Network Routing of Community Data Centers}\vspace*{-5pt}
\label{fig1}
\end{figure}

\vspace*{5pt}
The community data centers should be located near the main network entry point of the ISP gateway in the community. The servers can be installed in the same networking room as the ISP community switch location or in proximity to it. As shown in Figure \ref{fig1}, home Internet modems are connected to the ISP community gateway facility, which in turn connects to the nearby community data centers. 
\\\\
Within the community data centers, there are AI racks and gaming racks. Figure \ref{fig1} differentiates the routing of the network between gaming racks and AI racks because AI racks require much greater security. Specifically, the servers within AI racks will only be connected to the community customer intranet and not accessible from the broader Internet. They will also not store consumer private data after processing requests. However, the gaming racks are connected to both the community intranet and the broader internet to fulfill nearby community gaming demands if necessary.

\subsection{Remote Management}
The health of servers within community data centers will be remotely monitored by a centralized source. This is to minimize the presence of people on site for software problems. The Baseboard Management Controllers (BMCs) shown in Figure \ref{fig1} are managed remotely by the operators to monitor the status of the rack and server. Hardware problems will still need a technical person on site to repair and replace parts. Because the hardware is homogeneous, the training required to do this work should be very minimal compared to that of larger data centers.

\subsection{Tiered Cloud Resource}
The gaming racks in the community data center are located right at the center of the community networking gateway and can connect to the broader Internet. When community data centers are at capacity, they can use nearby data centers or reach hyperscale data centers for backup.

\subsection{Fiber to the Home / Fiber to the Room}
There is a wave to provide US homeowners with fiber internet. For example, AT\&T and Xfinity are pushing for the ultra-high-speed Internet using fiber technology\cite{ATTfiber}\cite{Xfinity}. This benefits the use case of community data centers. When the Internet speed of the user is not limited by the local transmission, the proximity to the datacenter becomes a bottleneck. In this case, community data centers will have much lower latency compared to larger data centers because of fewer networking hops to go through and proximity to the end users.

\section{Business Aspect of \\Community Data Centers}
\vspace*{5pt}
Community data centers are best partnered with Internet Service Providers (ISPs) due to their existing infrastructures. As an example, AT\&T currently sells the Direct TV, cellular, and fiber internet in bundles [18]. It would be easy to add gaming/AI into the existing bundle. In the process, ISP providers may collaborate with cloud providers to get support by renting servers and cloud solutions. The advantage of this is that cloud companies can deploy mature cloud technologies in the system and provide technical support directly to the average consumer with efficiency. In addition, after a few years of usage, the GPU servers may be outdated and need to be replaced. The rental option would make it easier to replace old units with the latest technology.
\\\\
In addition, community data centers will use homogeneous hardware. That will create an impact like budget airlines, where the maintenance and training costs of operating such servers are minimal.

\subsection{Optimized Speed for Community Data Center Connectivity}
A lot of times, ISPs limit their users' internet bandwidth because of the cost to reach broader Internet and not the local facility. ISP can allow maximized connection speed to connect to those local community data centers.

\subsection{Subscription Billing Model}
The subscription model proves lucrative for operators, as it provides consistent cash flow. In addition to that, the training of AI models would be a recurring activity because consumers will have new files that they want to sync with the existing AI model. Having a subscription model also gives consumers the flexibility of canceling the service when they move or no longer need it.

\section{Conclusion}
\vspace*{5pt}
Community data centers, different from traditional data centers or typical edge data centers, will be the new addition to the datacenter ecosystem in the near future. It brings low latency and quality computing resources to average consumers to meet increasingly large AI and gaming needs. In the technical design section, we showed how it uses state-of-the-art cloud technologies by using remote management and a tied cloud network to meet consumer demands. In the business section, we explored how such data centers can be implemented through a homogeneous hardware and subscription model.

\def\refname{References}

\vspace*{-8pt}

\begin{IEEEbiography}{Tianhao Zhang}{\,}is an independent researcher who completed his master's degree in Electrical and Computer Engineering at Cornell University in 2020. After graduation, he worked as a software engineer and served as the Nvidia A100 sustaining lead in Oracle's BMC embedded software team. In 2022, he transitioned to the OCI Server Management Group, where he currently holds the title of Senior Member of Technical Staff, working with Oracle’s CPU and GPU servers. This paper is part of his personal independent research and does not contain proprietary information from his current employer, Oracle. He can be contacted via Email at tz373@cornell.edu or via LinkedIn at www.linkedin.com/in/tianhao-zhang. \vadjust{\vfill\pagebreak} 
\end{IEEEbiography}

\end{document}